\documentclass{article}
\usepackage{arxiv}

\usepackage{amsmath}
\usepackage{graphicx}
\usepackage{url}
\usepackage{eurosym}
\usepackage{xspace}
\usepackage{textcomp}

\usepackage{filecontents}

\usepackage[nolist,nohyperlinks]{acronym}
\usepackage{cleveref}

\makeatletter
\newcommand*{\org@overidelabel}{}
\let\org@overridelabel\@verridelabel
\@ifpackagelater{acronym}{2015/03/21}{
  \renewcommand*{\@verridelabel}[1]{%
    \@bsphack
    \protected@write\@auxout{}{\string\AC@undonewlabel{#1@cref}}%
    \org@overridelabel{#1}%
    \@esphack
  }%
}{
  \renewcommand*{\@verridelabel}[1]{%
    \@bsphack
    \protected@write\@auxout{}{\string\undonewlabel{#1@cref}}%
    \org@overridelabel{#1}%
    \@esphack
  }%
}
\makeatother
\Crefformat{figure}{#2Fig.~#1#3}
\Crefmultiformat{figure}{Figs.~#2#1#3}{ and~#2#1#3}{, #2#1#3}{ and~#2#1#3}
\Crefformat{table}{#2Tab.~#1#3}
\Crefmultiformat{table}{Tabs.~#2#1#3}{ and~#2#1#3}{, #2#1#3}{ and~#2#1#3}
\Crefformat{equation}{#2Eq.~#1#3}
\Crefmultiformat{equation}{Eqs.~#2#1#3}{ and~#2#1#3}{, #2#1#3}{ and~#2#1#3}

\usepackage{tikz}
\usepackage{calc}
\usetikzlibrary{arrows,automata,matrix}
\usetikzlibrary{arrows,arrows.meta}
\usetikzlibrary{positioning,arrows}
\usetikzlibrary{shapes,arrows,fit,calc,positioning,automata}
\usepackage{float}

\let\OldTexttrademark\texttrademark
\renewcommand{\texttrademark}{\OldTexttrademark\xspace}%

\usepackage[nolist,nohyperlinks]{acronym}
\begin{acronym} 
    \acro{BMBF}{Federal Ministry of Education and Research}
    \acro{CPU}{Central Processing Unit}
    \acro{DoS}{Denial of Service}
    \acro{DMA}{Direct Memory Access}
    \acro{HAL}{Hardware Abstraction Layer}
    \acro{ICS}{Industrial Control System}
    \acrodefplural{ICS}{Industrial Control Systems}
    \acro{IDS}{Intrusion Detection System}
    \acro{IIoT}{Industrial Internet of Things}
    \acro{IO}{input/output}
    \acro{IoT}{Internet of Things}
    \acro{IP}{Internet Protocol}
    \acro{LwIP}{Lightweight IP}
    \acro{MCU}{Microcontroller Unit}
    \acro{PLC}{Programmable Logic Controller}
    \acrodefplural{PLC}{Programmable Logic Controllers}
    \acro{PoC}{Proof of Concept}
    \acro{RAM}{Random-Access Memory}
    \acro{ROM}{Read-only Memory}
    \acro{RTOS}{Real-time Operating System}
    \acro{SPI}{Serial Peripheral Interface}
    \acro{STM}{STMicroelectronics}
    \acro{SWD}{Serial Wire Debug}
    \acro{USB}{Universal Serial Bus}
\end{acronym}

\newcommand\copyrighttext{%
  \footnotesize \textcopyright 2019 IEEE. 
  Personal use of this material is permitted.
  Permission from IEEE must be obtained for all other uses,
  in any current or future media,
  including reprinting/republishing this material for advertising or promotional purposes,
  creating new collective works, for resale or redistribution to servers or lists,
  or reuse of any copyrighted component of this work in other works.
  DOI: 10.1109/MECO.2019.8760188 -- \url{http://dx.doi.org/10.1109/MECO.2019.8760188}
}
\newcommand\copyrightnotice{%
\begin{tikzpicture}[remember picture,overlay]
\node[anchor=south,yshift=10pt] at (current page.south) {\fbox{\parbox{\dimexpr\textwidth-\fboxsep-\fboxrule\relax}{\copyrighttext}}};
\end{tikzpicture}%
}

\title{A Secure Dual-MCU Architecture for Robust Communication of IIoT Devices}

\author{
Matthias Niedermaier \\
Hochschule Augsburg \\
matthias.niedermaier@hs-augsburg.de
\And
Dominik Merli \\
Hochschule Augsburg\\
dominik.merli@hs-augsburg.de
\And
Georg Sigl \\
TU M\"unchen\\
sigl@tum.de
}

\begin{document}

\maketitle

\begin{abstract}
The \ac{IIoT} has already become a part of our everyday life be it water supply, smart grid, or production, \ac{IIoT} is everywhere.
For example, factory operators want to know the current state of the production line.
These new demands for data acquisition in modern plants require industrial components to be able to communicate.
Nowadays, network communication in \acp{ICS} is often implemented via an \acs{IP}-based protocol.
This intercommunication also brings a larger attack surface for hackers.
If an \ac{IIoT} device is influenced by attackers, the physical process could be affected.
For example, a high network load could cause a high \ac{CPU} load and influence the reaction time on the physical control side.
In this paper, we introduce a dual \ac{MCU} setup to ensure a resilient controlling for \ac{IIoT} devices like \acp{PLC}.
We introduce a possible solution for the demand of secure architectures in the \ac{IIoT}.
Moreover, we provide a \ac{PoC} implementation with a benchmark and a comparison with a standard \ac{PLC}.
\end{abstract}
\keywords{security, robust, iiot, industrial control system}
\copyrightnotice
\acresetall

\section{Introduction}
New demands for data collection in industrial plants call for a higher degree of networking.
This partly dissolves the historical air gap segmentation between office and operational networks.
In modern systems, it is possible to have access to the \ac{IIoT}, e.g.
through the company network.
Owing to this development in the \ac{IIoT}, the attack surface is also getting larger.
This will require that the individual components must be secure by design to reduce the probability and impact of successful attacks.
Considering a \ac{PLC} as an example of an \ac{IIoT} device, these have a special meaning, because they mostly control a physical process.
This means that influencing a \ac{PLC} could also influence the process that takes place in the real world.

Previous research has already shown that \acp{PLC} could be affected by network traffic, e.g. through cyber-attacks.
One of the first was presented by Dillon Beresford on Siemens S7 \acp{PLC}~\cite{beresford2011exploiting}, where a simple message could start and stop these \acp{PLC}.
This was later demonstrated on other devices with different protocols and vulnerabilities (e.g.
Phoenix Contact~\cite{niedermaier2017propfuzz} and Beckhoff~\cite{bonney2015ics}).

Another possibility to influence the cycle time behavior of a \ac{PLC} is network flooding~\cite{niedermaier18woot}.
In this case, the \ac{PLC} is busy processing the network data and is not able to control the physical process in the expected way.

Most existing \ac{IIoT} devices have one \ac{MCU} that handles the network and \ac{IO} operations.
If this microcontroller can be influenced, then the behavior of the physical inputs and outputs are also affected.
For the real world, this means that the physical process controlled by the \ac{PLC} is vulnerable.

As a result, secure architectures for \ac{IIoT} devices are necessary~\cite{cardenas2008research},~\cite{cardenas2008secure}.
The \ac{IIoT} architecture comprising two \acp{MCU}, a network \ac{MCU} (NW-\ac{MCU}), and an \ac{IO}-\ac{MCU} handling the connection to the sensors and actuators, presented in this paper, offers the following advantages compared to most existing solutions:
\begin{itemize}
    \item A well-controlled communication between the two microcontrollers reduces intentional and unintentional influencing of the physical process.
    \item Compared to a software solution, e.g. based on a single \ac{MCU} and an \ac{RTOS}~\cite{nguyen2015design}, a vulnerability in the hardware or software of the network \ac{MCU} will not directly influence the \ac{IO} \ac{MCU}.
    \item The reduced code size on the \ac{IO} \ac{MCU} reduces testing effort, e.g. for safety certifications, because the critical code size is smaller.
    \item By using a unidirectional connection~\cite{zilberstein2010protection} from the \ac{IO} \ac{MCU} to the network \ac{MCU}, it is possible to monitor without influencing.
\end{itemize}

The paper is structured as follows.
\Cref{sec:methodology} explains the methodology behind a dual controller setup for secure controlling and gives the necessary background.
In \Cref{sec:implementation}, the \ac{PoC} implementation is illustrated.
\Cref{sec:benchmarking} compares the robustness against \acl{DoS} attacks of the secure architecture and a commercial PLC.
Finally, a conclusion is given in \Cref{sec:conclusion}.

\section{Methodology and Background\label{sec:methodology}}
One of the simplest attacks in networks is a \acf{DoS} flooding attack~\cite{schuba1997analysis}.
This means that as many packets as possible are sent into a network, to either interfere with the network communication or to generate a high CPU load on the target in such a way that other programs cannot be processed properly.
The effects are especially dangerous for \ac{IIoT} devices such as \acp{PLC}, as they interact with the physical world, resulting in serious damage or injury.
Haddadin et al. showed how strong robots that interact with humans can injure them~\cite{haddadin2007safety}.
This cannot only happen because of a bug in the program, but also due to \ac{DoS} attacks over the network.
Therefore, the communication part of \acp{PLC} should be designed in such a way that it does not influence the control part.

There are secure architectures for complete chips with patents available on the 
market~\cite{fruehling2005secured}.
This concept requires deep knowledge and no standard \ac{MCU} can be used, which may make the end product expensive.
Alves et al. introduced an open-source Linux-based \ac{PLC} and implemented an \ac{IDS} as a \ac{DoS} protection~\cite{alves2018embedding}.
However, the used method only partially protects against \ac{DoS} attacks, implementation errors, and zero day vulnerabilities.
Furthermore, it is currently only feasible for Linux-based systems.
They also show that the program execution time on current \acp{PLC} varies during their tests.

\begin{figure}[htb!]
    \centering
    \begin{tikzpicture}[node distance=1cm,
        auto,
        block/.style={
            rectangle,
        draw=black,
        align=center,
        rounded corners,
        dashed
    }
        ]
        \coordinate (a) at (1.1,1);
        \coordinate (b) at (5.2,1);

        \draw [-, dashed, color=red!80!black, line width=0.5mm] (5.2,1.7) to (5.2,-0.4);
        \node[align=center, minimum width=1.5cm, minimum height=0.9cm, anchor=west,
        text width=2.5cm, color=red!80!black] at (5.3,0) (x) {\small Critical Control \\ \small Part};
        \node[align=center, minimum width=1.5cm, minimum height=0.9cm, anchor=west,
        text width=1.6cm, color=green!40!black] at (2,0) (x) {\small Uncritical \\ \small Part};

        \node[rectangle, draw, align=center, minimum width=1.5cm, minimum height=0.9cm, anchor=west,
        text width=1.6cm] at (a) (m) {\small MCU \\ \small Network};
        \node[rectangle, draw, align=center, minimum width=1.5cm, minimum height=0.9cm, anchor=west,
        text width=1.6cm, fill=white] at (b) (n) {\small MCU \\ \small Control/IO};

        \draw [{Stealth[scale=1.0]}-{Stealth[scale=1.0]}]  (m.east) to node[above] {\small Robust}
        node[below] {\small communication} (n.west);	

        \draw [-] ([xshift=0.0cm, yshift=0.4cm]n.east) to node[above] {\small IO} 
        ([xshift=1.0cm, yshift=0.4cm]n.east);	
        \draw [-] ([xshift=0.0cm, yshift=-0.0cm]n.east) to node[above] {\small IO} 
        ([xshift=1.0cm, yshift=-0.0cm]n.east);	
        \draw [-] ([xshift=0.0cm, yshift=-0.4cm]n.east) to node[above] {\small IO} 
        ([xshift=1.0cm, yshift=-0.4cm]n.east);

        \draw [-{Stealth[scale=1.0]}]  ([xshift=-1.4cm, yshift=-0.0cm]m.west) to 
        node[above, minimum width=1.5cm, text width=1.4cm] {\small Network} (m.west);

    \end{tikzpicture}
    \caption{Example architecture of a dual \ac{MCU} set-up for robust controlling.}
    \label{fig_exparchitecture}
\end{figure}
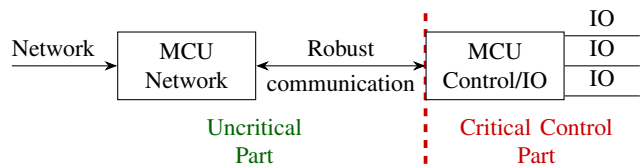

\Cref{fig_exparchitecture} shows the principle of our secure architecture introduced in this paper with a dedicated network and \ac{IO} \ac{MCU}.
The network \ac{MCU} handles the communication over Ethernet and Modbus/TCP.
Configuration information and control data are submitted without influencing the \ac{IO} \ac{MCU}.
For this to happen, the \ac{IO} control must be handled in a predefined time slice by the \ac{IO} \ac{MCU} to ensure a certain response time.
The uncritical part is responsible for the network communication and the critical control part, for the physical process.
Influences on this critical part will affect the real world.

The cycle time of a \ac{PLC} is the time for the execution of a complete program cycle including the communication.
It depends on the processing time of the program, which is determined by the number of instructions.
Higher prioritized tasks interrupt the cycle, thereby delaying the actual cycle.
This cycle time ($t_{cycle}$) is the sum of the phases that are processed at each pass.
In a simplified representation with one task, there are four phases.
First of all, the inputs get read in ($t_{read\_in}$); this step has a constant processing time.
It is independent of input changes for cyclic tasks.
Thereafter, the communication is handled ($t_{comm}$).
The communication depends on external participants and can have different runtimes.
For example, if the bus speed is slow or the data size is big, the communication part takes longer.
At the end, the necessary calculation ($t_{calc}$) is done and the outputs are written back ($t_{write\_out}$).
In this case (\Cref{eq_cycle}), the cycle time ($t_{cycle}$) is free-running and varies in time.

\begin{equation} \label{eq_cycle}
    t_{cycle} = t_{read\_in} + t_{comm} + t_{calc} + t_{write\_out}
\end{equation}

For our approach, the \ac{IO} \ac{MCU} must have a constant runtime independent of the network \ac{MCU}, which results in the requirement of a constant cycle time ($t_{cycle}$).
This could be achieved by setting a timeout to the communication between the network \ac{MCU} and the \ac{IO} \ac{MCU}.
To get a constant cycle time ($t_{cycle}$) of the \ac{IO} \ac{MCU}, a delay ($t_{delay}$) is inserted to equalize time fluctuations.
The calculation of the delay is shown in \Cref{eq_delay}.

\begin{equation} \label{eq_delay}
    t_{delay} = t_{cycle} - (t_{read\_in} + t_{comm} + t_{calc} + t_{write\_out})
\end{equation}

It must be ensured that the cycle time ($t_{cycle}$) is higher than the maximum time, which can pass through the four phases in \Cref{eq_cycle}.
Therefore, the maximum time of each phase must be limited depending on the desired cycle time.
The behavior, which is illustrated in \Cref{eq_delay}, must be represented by the \ac{IO} \ac{MCU} and runs independent of the NW \ac{MCU}.

\section{\ac{PoC} Implementation \label{sec:implementation}}
To prove the feasibility of the introduced method, a \ac{PoC} implementation is necessary.
The focus is on the robust communication between the two \acp{MCU} and the real-time behavior of the \ac{IO} control during flooding attacks.

\subsection{\ac{PoC} Hardware}
The hardware of the secure architecture consists of the network board and the \ac{IO} shield, which are connected.
\Cref{tab_boardinfo} shows the specification of the hardware used.
The \ac{MCU} on the network board is faster but more expensive than the \ac{IO} \ac{MCU}.

\begin{table}[htb]
    \centering
    \caption{Specification of the used hardware for the \ac{PoC} implementation.}
    \label{tab_boardinfo}
        \begin{tabular}{l l l}
            \hline
            \textbf{Hardware} & \textbf{Network Board} & \textbf{\ac{IO} Shield} \\
            \hline
            Board design      & STMicroelectronics     & custom         \\
            \ac{MCU}          & STM32F767ZIT6          & STM32F030F4P6  \\
            Core              &
            ARM\textsuperscript{\tiny{\textregistered}}Cortex\textsuperscript{\tiny{\textregistered}}-M7      
                              &
            Arm\textsuperscript{\tiny{\textregistered}}Cortex\textsuperscript{\tiny{\textregistered}}-M0 \\
            Clock             & up to 216 MHz          & up to 48 MHz   \\
            \acs{RAM}         & 512kB                  & 4kB            \\
            Flash             & 2MB                    & 16kB           \\
            \ac{MCU} price    & $\sim$ 10\euro         & $\sim$ 1\euro  \\
        \end{tabular}
\end{table}

For \acp{PLC}, which often costs several hundred dollars, an additional \ac{IO} \ac{MCU} would not render the final product much more expensive.
Furthermore, the proposed concept is possible with different \acp{MCU}, depending on the later demands.

\subsubsection{Network Board Hardware}
For the network \ac{MCU}, a development board (STM NUCLEO-F767ZI\footnote{\url{https://www.st.com/en/evaluation-tools/nucleo-f767zi.html}})
is used.
This \ac{MCU} was chosen because, on the one hand, this series is relatively energy-efficient, which is also used in standard \acp{PLC}, and on the other, offers enough performance for further evaluations.
The board provides an RJ45 Ethernet connector and an Arudino\texttrademark Uno V3 connector.
Additionally, an ST-Link programmer with \ac{SWD} and serial communication is attached to the \ac{MCU} for programming and debugging output.

\subsubsection{\ac{IO} Shield Hardware}
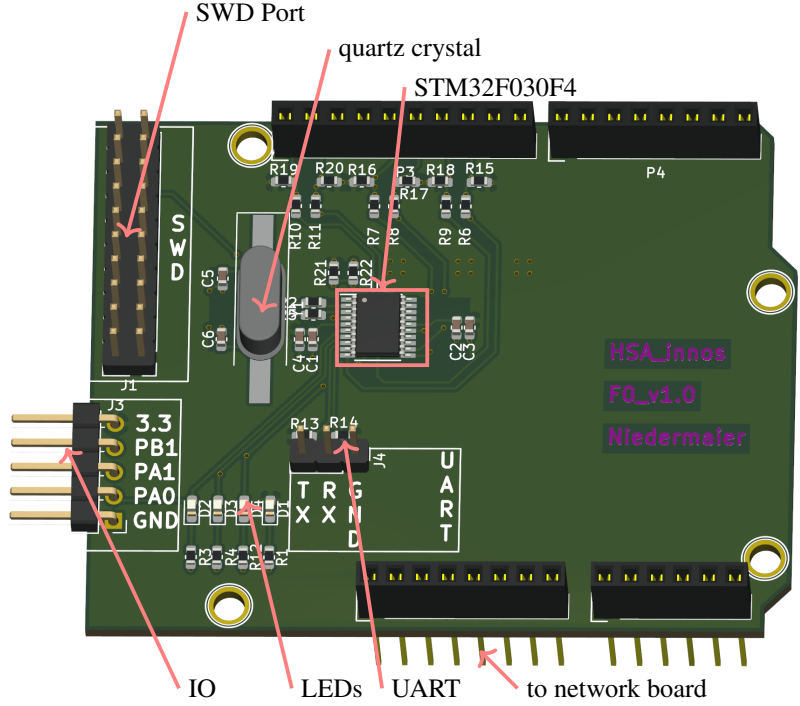
\begin{figure}[htb]
    \centering
    \begin{tikzpicture}
        \node[inner sep=0pt] (shield) at (0,0)
            {\includegraphics[width=.65\columnwidth]{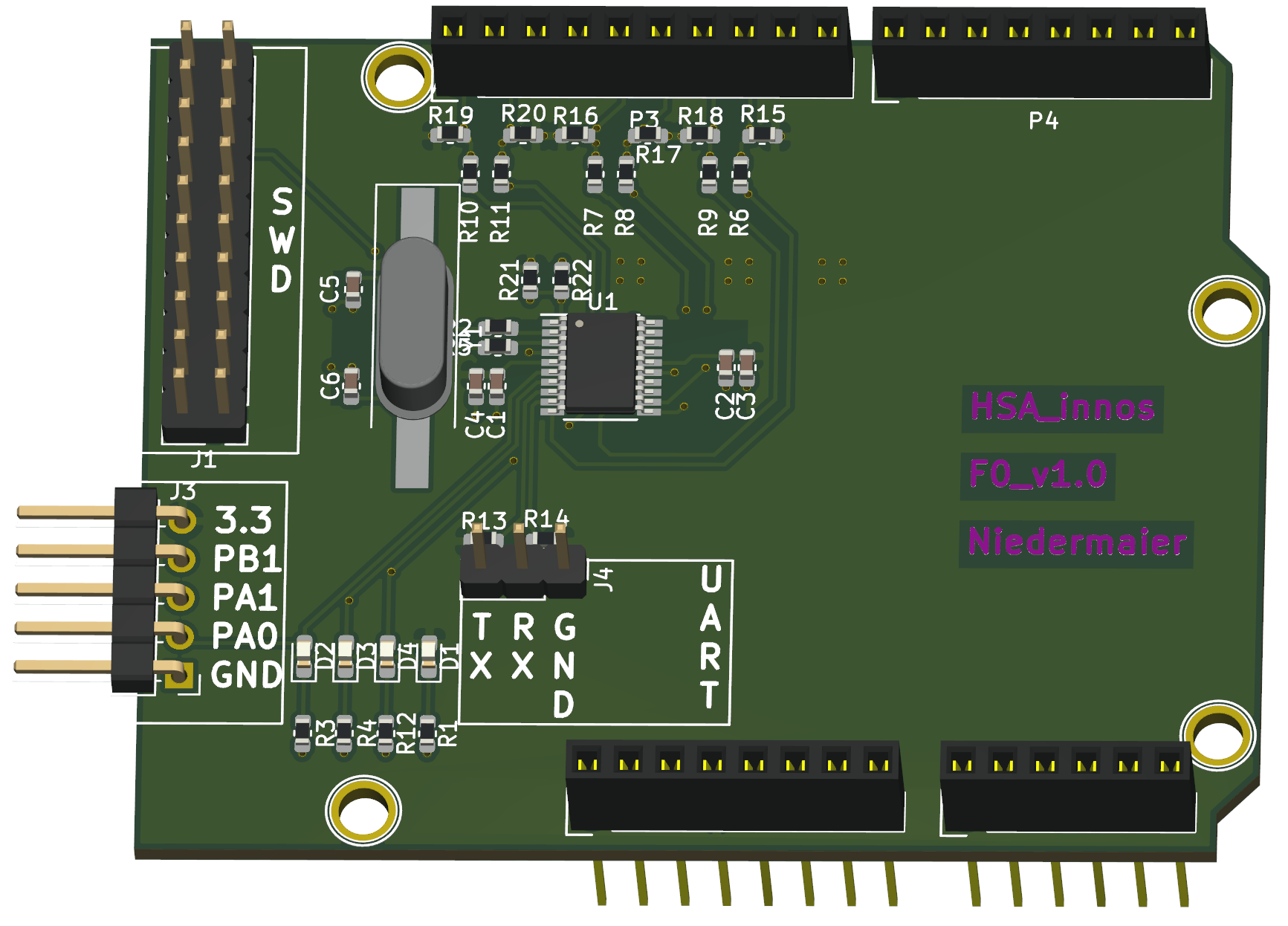}};
        \draw[color=red!50, line width=0.5mm] (-0.9,0.3) -- (0.3,0.3) -- (0.3,1.3) -- (-0.9,1.3) -- cycle;
        \node[align=left,anchor=west] at (0,4) {STM32F030F4};
        \draw[->,color=red!50, line width=0.5mm] (0,4) -- (-0.3,1.3);
        \node[align=left,anchor=west] at (-2.9,5.0) {SWD Port};
        \draw[->,color=red!50, line width=0.5mm] (-2.9,5.0) -- (-3.7,2);
        \node[align=left,anchor=west] at (-1.0,4.5) {quartz crystal};
        \draw[->,color=red!50, line width=0.5mm] (-1.0,4.5) -- (-1.9,1.0);
        \node[align=left,anchor=west] at (-3.0,-4) {\ac{IO}};
        \draw[->,color=red!50, line width=0.5mm] (-3.0,-4) -- (-4.5,-0.8);
        \node[align=left,anchor=west] at (-1.5,-4) {LEDs};
        \draw[->,color=red!50, line width=0.5mm] (-1.5,-4) -- (-2.1,-1.5);
        \node[align=left,anchor=west] at (1.5,-4) {to network board};
        \draw[->,color=red!50, line width=0.5mm] (1.5,-4) -- (1.0,-3.5);
        \node[align=left,anchor=west] at (-0.3,-4) {UART};
        \draw[->,color=red!50, line width=0.5mm] (-0.3,-4) -- (-0.8,-0.7);
    \end{tikzpicture}
    \caption{Rendered controller shield that is placed on top of the network \ac{MCU} board.}
    \label{fig_pcb}
\end{figure}

The \ac{IO} shield is a custom design, as there is no suitable shield for this purpose.
\Cref{fig_pcb} shows the shield, which is designed to be compatible with the Arduino\texttrademark Uno V3 header.
This makes the shield usable with many other boards.

\subsection{\ac{PoC} Software}
As explained in \Cref{sec:methodology}, the communication part between the two \acp{MCU} does not use a constant time during processing.
This requires calculation and compensation to achieve a uniform runtime, resulting in a constant cycle time.
The software that runs on the two \acp{MCU} is fundamentally different in terms of \ac{RAM} and \ac{ROM} usage.
Furthermore, in contrast to the \ac{IO} \ac{MCU}, the network \ac{MCU} has no ``hard'' real-time requirements.
Of course, this only applies if the network communication does not have real-time requirements.

\subsubsection{Network \ac{MCU}}
\begin{figure}[htb]
    \centering
    \begin{tikzpicture}
        \node[inner sep=0pt] (webside) at (0,0)
            {\includegraphics[width=.45\columnwidth]{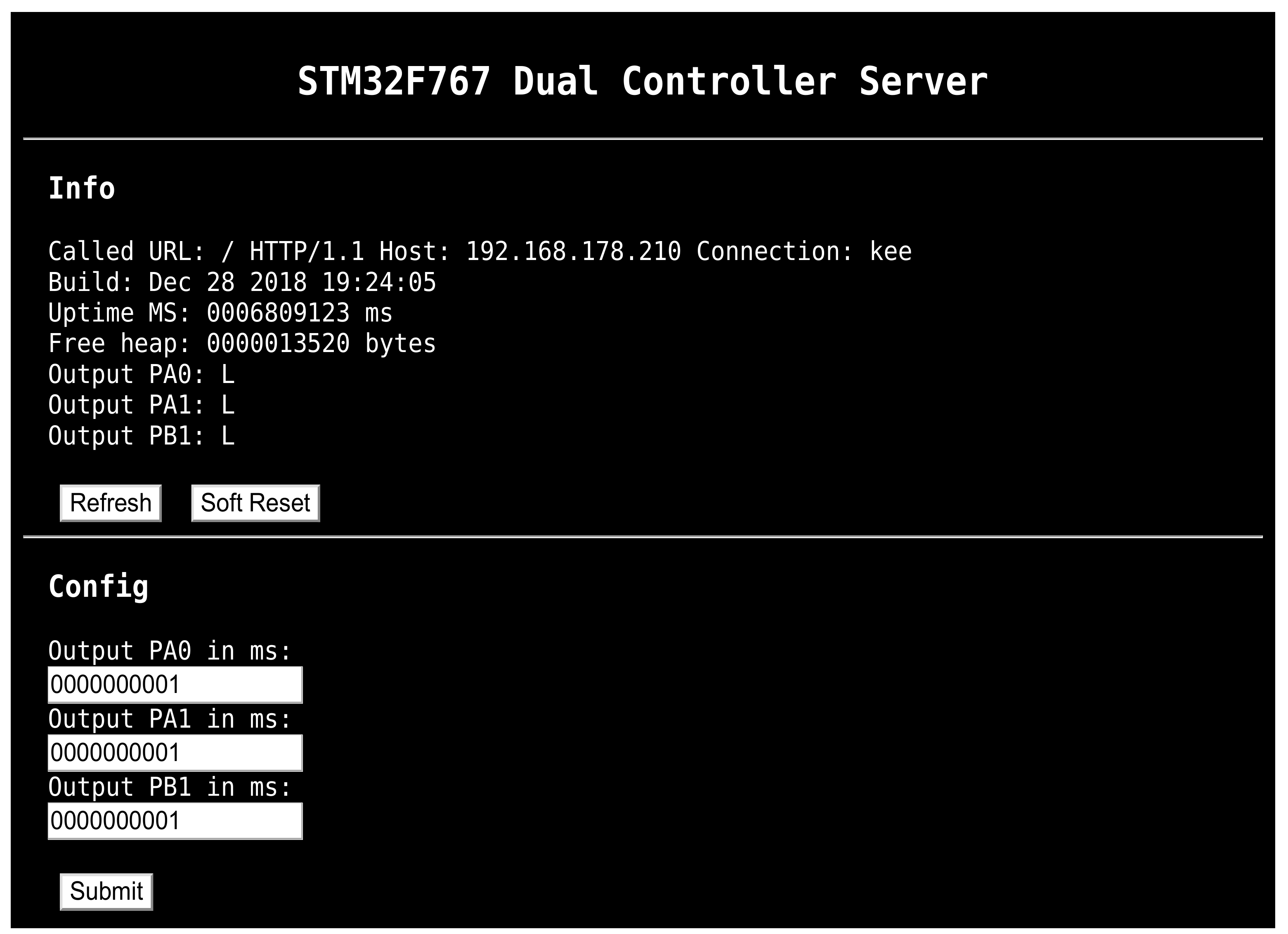}};
    \end{tikzpicture}
    \caption{Website, running on the network \ac{MCU}, showing some information and allowing configuration of the network and the \ac{IO} \ac{MCU}.}
    \label{fig_webside}
\end{figure}

The network board has a much higher computing power than the \ac{IO} board.
It runs an operating system (FreeRTOS\footnote{https://www.freertos.org/}) to handle the different tasks in a pre-emptive multi-tasking single-core implementation.
As a result, the network communication with multiple subscribers can be handled through different \ac{RTOS} tasks.
For the network communication, the \ac{LwIP}\footnote{https://savannah.nongnu.org/projects/lwip/} stack is used.
\Cref{fig_webside} shows the configuration web server running on the network board.
This shows actual information, such as the uptime and the current state of the outputs.
Furthermore, the cycle time of the outputs on the \ac{IO} \ac{MCU} can be configured.

For the communication between the two boards, an \ac{SPI} with a speed of 13.5Mbit/s is used.
The network \ac{MCU} is the master and continuously transmits the information to the \ac{IO} \ac{MCU}.
If the \ac{IO} \ac{MCU} does not respond within a certain time, the transmission is tried again after a delay.

\subsubsection{IO \ac{MCU}}
\begin{figure}[htb]
    \centering
    \footnotesize
    \begin{tikzpicture}[>=stealth',shorten >=2pt,auto,shorten <=2pt,node distance=0.6 cm,
        every state/.style={align=center,minimum size=1.8cm},
        every edge/.style={draw,thick},
        loop label/.style={draw,align=center,text width=2cm,outer sep=4pt,minimum height=1cm}
        ]

        \node[initial,state] (A)                {Initialize};
        \node[state]         (B) [right= of A]  {Read inputs \\$t_{read\_in}$};
        \node[state]         (C) [right= of B]  {Communi- \\cation \\$t_{comm}$};
        \node[state]         (D) [below= of C]  {Calculation \\$t_{calc}$};
        \node[state,dashed]  (E) [left= of D]  {Wait time \\$t_{delay}$};
        \node[state]         (F) [left= of E]  {Update \\outputs \\$t_{write\_out}$};

        \node[rotate=45] at (1.0,-1.0) {reset timer};

        \draw[->] ([xshift=0ex]A.east) -- ([xshift=0ex]B.west);
        \draw[->] ([xshift=0ex]B.east) -- ([xshift=0ex]C.west);
        \draw[->] ([xshift=0ex]C.south) -- ([xshift=0ex]D.north);
        \draw[->] ([xshift=0ex]D.west) -- ([xshift=0ex]E.east);
        \draw[->] ([xshift=0ex]E.west) -- ([xshift=0ex]F.east);
        \draw[->] ([xshift=0ex]F.north east) -- ([xshift=0ex]B.south west);

    \end{tikzpicture}
    \caption{Program sequence to have a defined time behavior of the \ac{IO} \ac{MCU}.
The dashed circle ``Wait time'' is the additional task compared to a standard \ac{PLC} cycle.}
    \label{fig_states}
\end{figure}
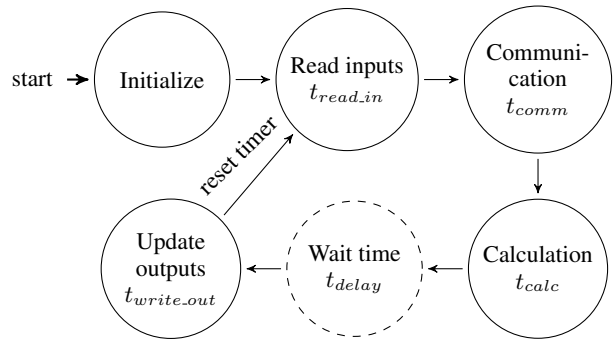

The \ac{IO} \ac{MCU} runs a bare metal system, with the usage of the \ac{STM} \ac{HAL}.
This makes later changes to the \ac{MCU} easier if, for example, more performance is necessary.
Within this \ac{HAL}, the SysTick is set to 100$\mu s$, which is also the resolution of all time-based \ac{HAL} functions, such as the timeouts of the communication functions.
The sequence of the program on the \ac{IO} \ac{MCU} is illustrated in \Cref{fig_states}.
The start represents the initial powering of the \ac{MCU}, whereon the initialization of this is done.

After the initialization, a continuous cycle is executed.
At the start of the cycle, the timer to measure the delay is reset.
After this, the inputs are read with the \ac{HAL} functions and the \ac{SPI} with a timeout of 500$\mu s$ is executed.
The time is chosen so that there is enough time to transfer the necessary data and a cycle time of 1$ms$ is possible.
After this, the calculation of the new output states is done by comparing the current cycle count with the configured cycle time.
In these cycles, the varying timing which is measured with the timer must be compensated (See \Cref{eq_delay}) to get a constant cycle time.
This is done in the wait state by holding it there until the desired cycle time is reached and then writing the previous calculated outputs back.
In the \ac{PoC} implementation, the cycle time is set to 1$ms$.
This is a common minimum cycle time for commercial \ac{PLC} solutions.
Owing to this, multiples of 1$ms$ can be used as the \ac{IO} response time.
This is common for current commercial \acp{PLC}.

To ensure robust communication over \ac{SPI}, the receive and transmit functions on the \ac{IO} are implemented in a blocking mode with a timeout.
Interrupts and \ac{DMA} are not used to prevent blocking and timing problems through many interrupts and memory overflows by overwriting buffer boundaries.
Thus, the \ac{IO} \ac{MCU} could miss a transmission from the NW \ac{MCU} to fulfill the real-time requirements of the \ac{IO} control.

\section{Benchmarking \label{sec:benchmarking}}
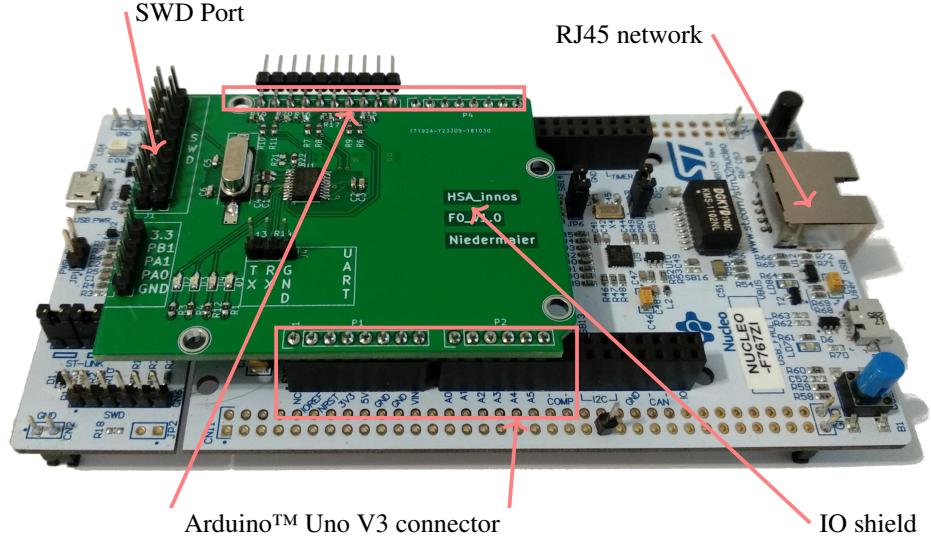
\begin{figure}[htb!]
    \centering
    \begin{tikzpicture}
        \node[inner sep=0pt] (setup) at (0,0)
            {\includegraphics[width=.75\columnwidth]{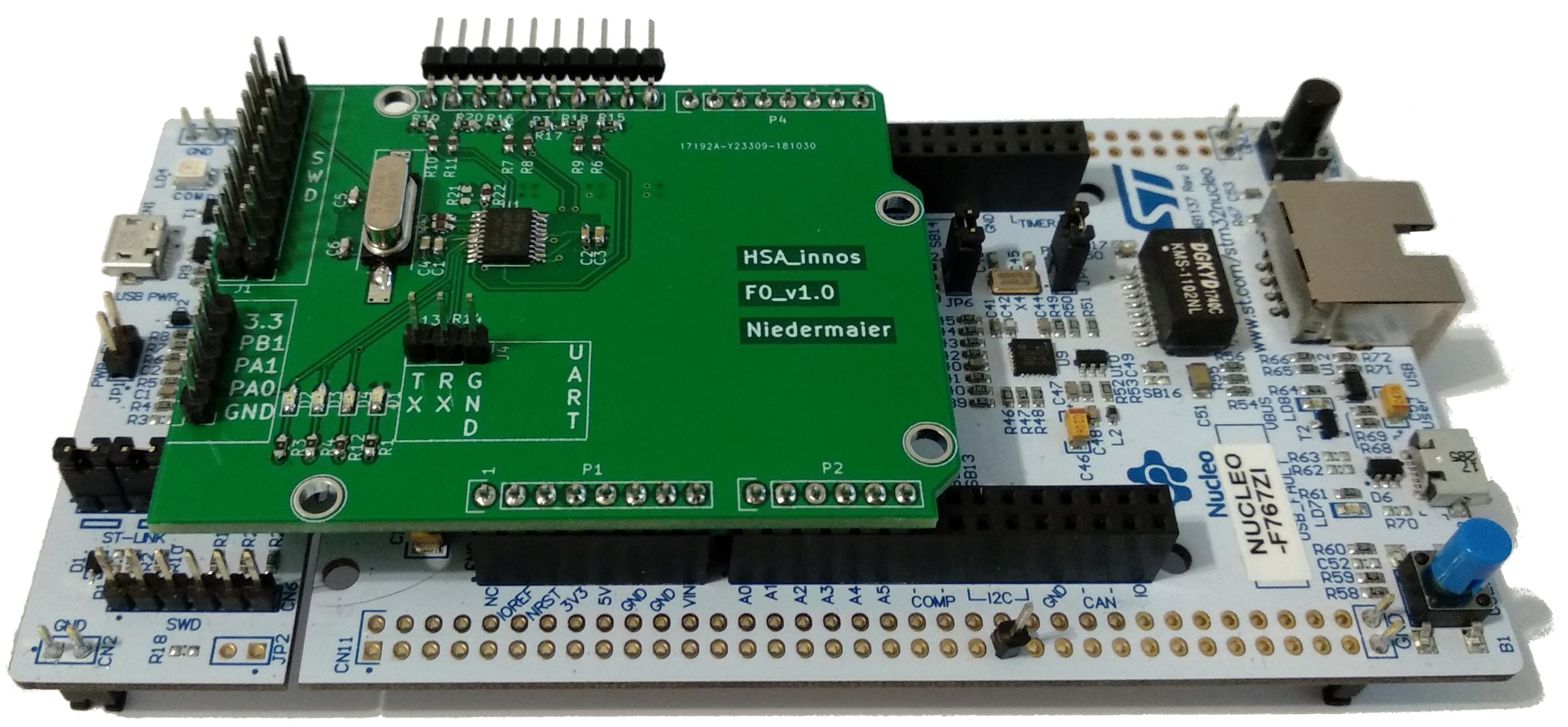}};
        \node[align=left,anchor=east] at (3.2,3) {RJ45 network};              
        \draw[->,color=red!50, line width=0.5mm] (3.2,3) -- (4.5,0.7);
        \node[align=left,anchor=west] at (4.5,-3.5) {\ac{IO} shield};              
        \draw[->,color=red!50, line width=0.5mm] (4.5,-3.5) -- (0.0,0.7);
        \node[align=left,anchor=west] at (-4.6,3.3) {SWD Port};              
        \draw[->,color=red!50, line width=0.5mm] (-4.6,3.3) -- (-4.2,1.4);
        \draw[color=red!50, line width=0.5mm] (-2.6,-2.1) -- (1.4,-2.1) -- (1.4,-0.9) -- (-2.6,-0.9) -- cycle;
        \node[align=left,anchor=east] at (0.5,-3.5) {Arduino\texttrademark  Uno V3 connector};              
        \draw[->,color=red!50, line width=0.5mm] (0.5,-3.3) -- (0.6,-2.1);
        \draw[color=red!50, line width=0.5mm] (-3.3,2.0) -- (0.7,2.0) -- (0.7,2.3) -- (-3.3,2.3) -- cycle;
        \draw[->,color=red!50, line width=0.5mm] (-4,-3.3) -- (-1.6,2.0);
    \end{tikzpicture}
    \caption{Image showing the complete setup with the network \ac{MCU} board and the \ac{IO} shield.}
    \label{fig_setup}
\end{figure}

\Cref{fig_setup} shows the \ac{PoC} setup with the network board and the attached \ac{IO} shield, which is used for the benchmark.
For a secure operation, the interferences on the \ac{SPI} bus must be considered so that the \ac{IO} \ac{MCU} is not influenced.
These include flooding by the network \ac{MCU}, invalid data, and shortcuts.
The current \ac{PoC} is not protected by cryptographic mechanisms.
Nevertheless, the \ac{SPI} communication cannot influence the \ac{IO} \ac{MCU}.

\begin{figure}[htb!]
    \centering
    \begin{tikzpicture}
        \node[inner sep=0pt] (time) at (0,0)
            {\includegraphics[width=.75\textwidth]{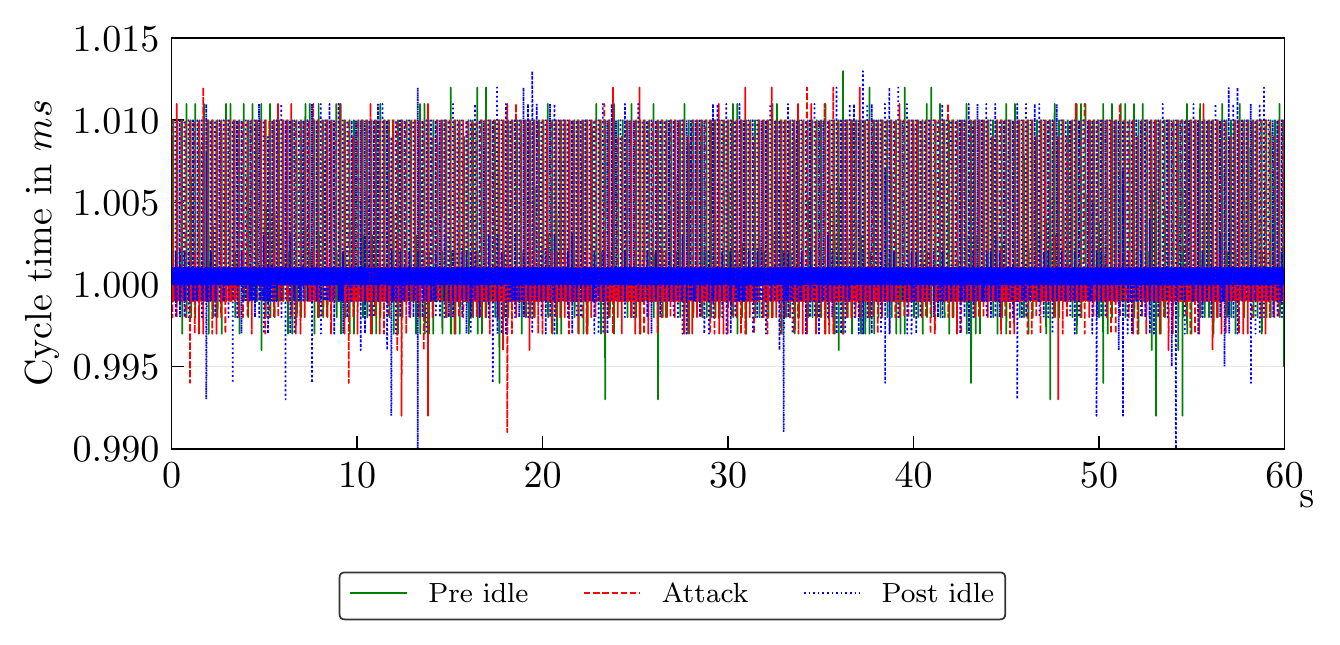}};
    \end{tikzpicture}
    \caption{Time plot of the 1ms cycle time during pre-idle, attack, and post-idle of our secure implementation.
The jitter is about 10$\mu s$, which is equal to a deviation of 1\%.}
    \label{fig_time_1ms}
\end{figure}

Additionally, to show the stability of the concept during network flooding attacks, the cycle time is measured during a flooding attack.
For the measurements, a PicoScope 2208B \ac{USB} oscilloscope is used.
With this, the measured data can be exported and analyzed.
\Cref{fig_time_1ms} shows the cycle time of our implementation over time during pre-idle,
hping3\footnote{\url{https://www.spirent.com/}} flooding attack, and post-idle.
The jitter is only about 10$\mu s$, which is equivalent to 1\% deviation.
The cycle time is similar in all phases and is not influenced by the attack.

Furthermore, our implementation is compared with a standard \ac{PLC}.
As this reference, a Wago \ac{PLC} (HW:750-8100 SW:02.05.23(08)) is used, which is a current \ac{PLC} from this vendor.
This should not be regarded as an opinion about this product, but as a reference for comparison.

\begin{figure}[htb]
    \centering
        \includegraphics[width=0.75\textwidth]{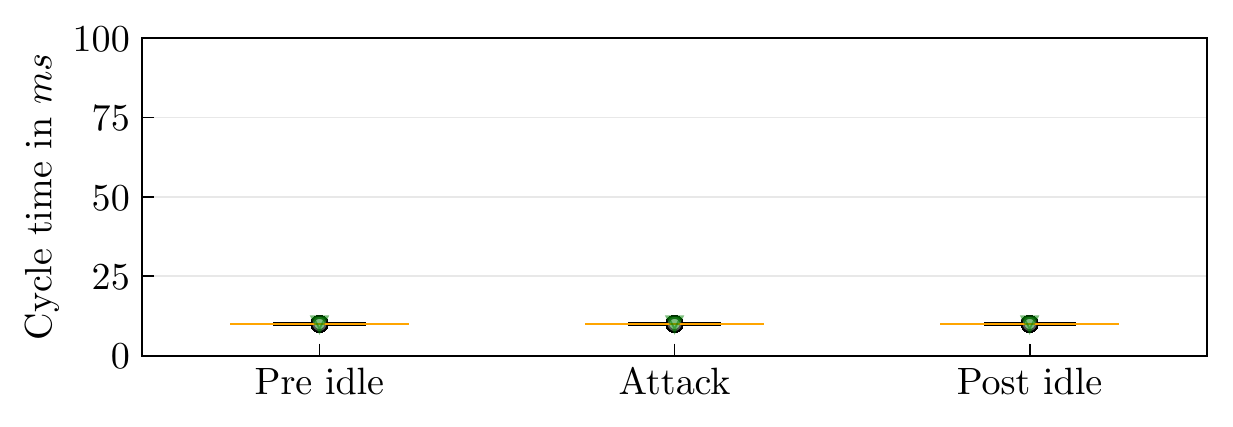}
        \caption{Boxplot of the cycle time of our approach during hping3 attack.
A constant cycle time is set to $10ms$ during all phases.}
        \label{fig:secure_box}
\end{figure}

\begin{figure}[htb]
     \centering
        \includegraphics[width=0.75\textwidth]{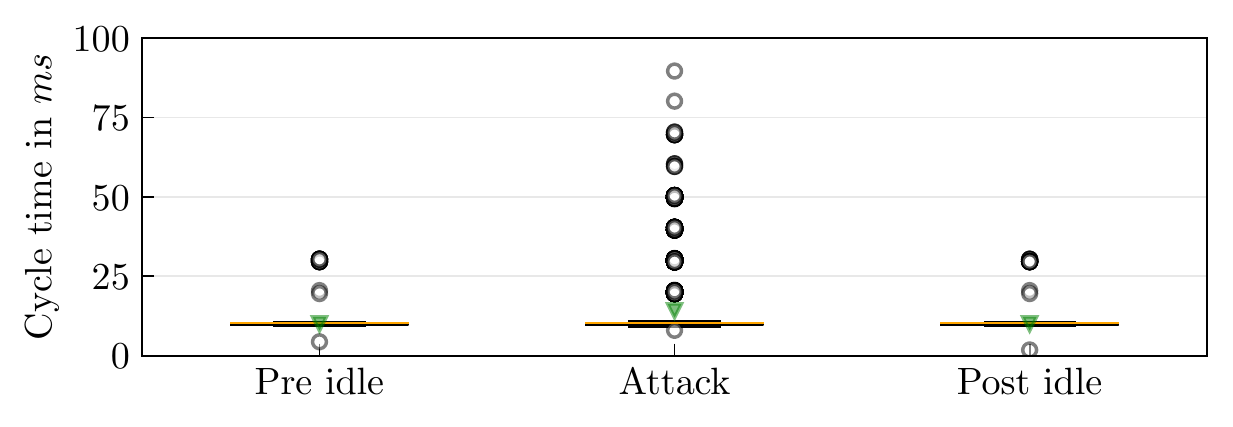}
        \caption{Boxplot of cycle time of the Wago \ac{PLC} during hping3 attack, with
        variances during idle and influences during attack.}
        \label{fig:wago_box}
\end{figure}

\begin{figure}[htb]
    \centering
        \includegraphics[width=0.75\textwidth]{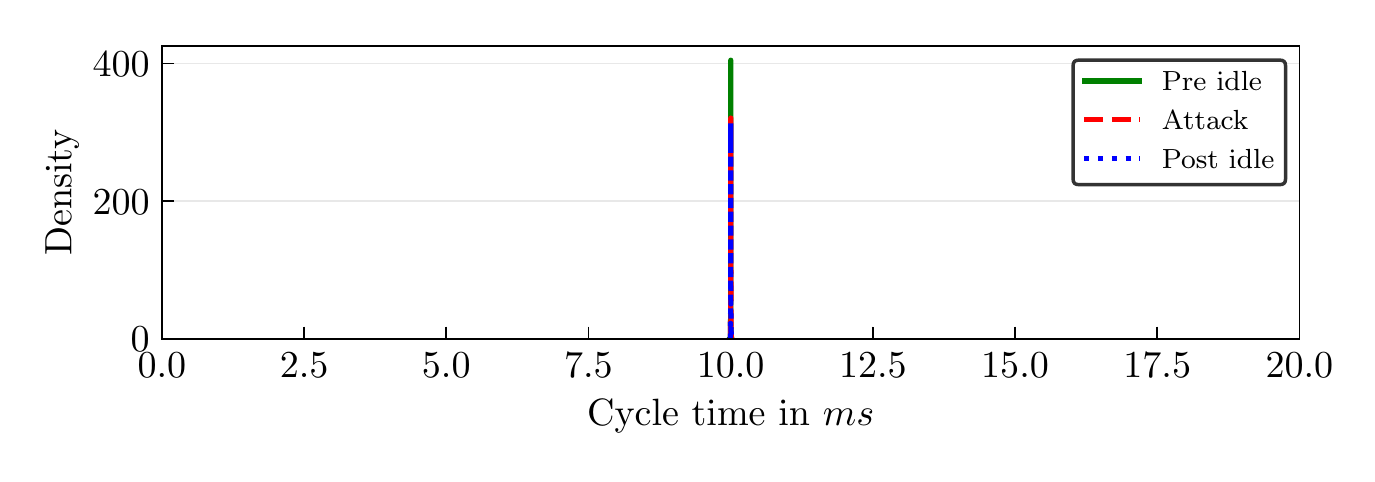}
        \caption{Density plot of the 10$ms$ cycle time of our implementation during hping3 attack.}
        \label{fig:secure_den}
\end{figure}

\begin{figure}[htb]
    \centering
        \includegraphics[width=0.75\textwidth]{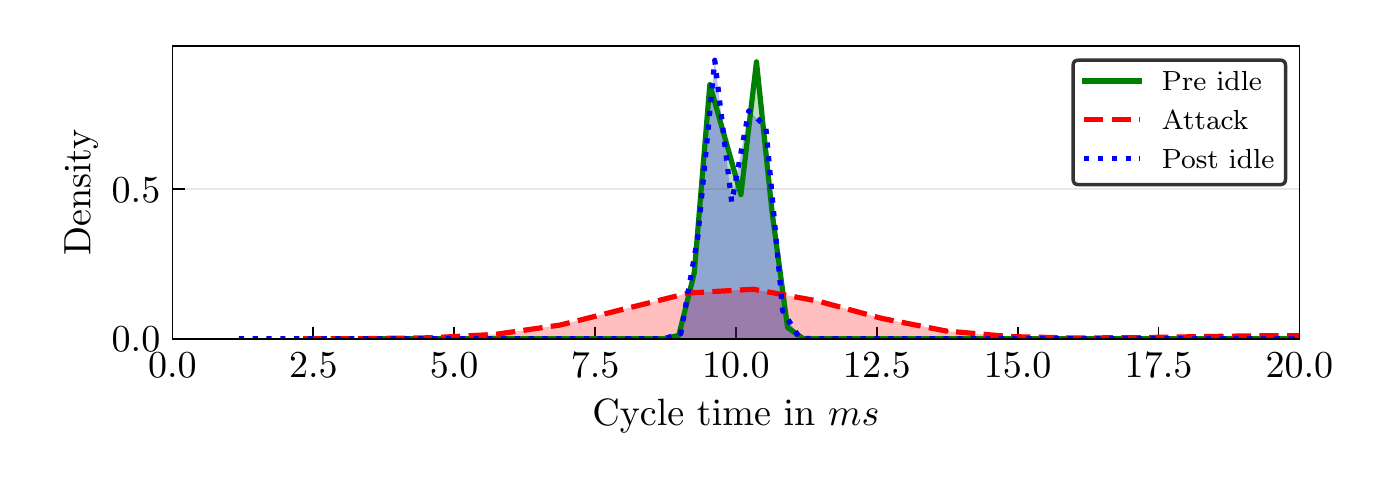}
        \caption{Density plot of the cycle time of the Wago \ac{PLC} during hping3 attack influenced during the attack.}
        \label{fig:wago_den}
\end{figure}

\Cref{fig:secure_box} shows the boxplot of our secure \ac{PoC} implementation during pre-idle, attack, and post-idle.
The measurement duration of each phase is 60$s$.
The \ac{PoC} introduced in this paper has a fixed cycle time of 1$ms$ and the \ac{PLC} from Wago has a default of 10$ms$.
For this reason, our implementation toggles the output every 10 cycles to allow a direct comparison.
The maximum jitter during idle is less than 1\% for the secure architecture and about 300\% for the Wago \ac{PLC}.
The \ac{DoS} attack is done with hping3 in the flooding mode.
No difference was observed in the cycle time on our \ac{PoC} implementation during pre-idle, attack, and post idle.
On the common \ac{PLC} (\Cref{fig:wago_box}), the attack slows down the cycle time noticeably~\cite{niedermaier18woot}.
The density representation (\Cref{fig:secure_den}) shows that our \ac{PoC} implementation is stable and only varies in a small range.
In contrast, \Cref{fig:wago_den} shows that on a common controller a flooding attack could influence the cycle time.
In this case, it ranges up to a cycle time of 100$ms$ (Factor 10 slower), where the outputs of the \ac{PLC} are not updated.
The comparison between our secure architecture presented here and a current \ac{PLC} shows that our proposed solution is feasible and stable.

\section{Conclusion \label{sec:conclusion}}
The presented architecture allows a secure and robust operation of an \ac{IIoT} device in a network environment.
This is achieved by a dual \ac{MCU} architecture where one takes over the hard timing requirements and the second controller handles the network communication.
This ensures that even with weak points in the software implementation, e.g. vulnerabilities in the network stacks or the operating system, the physical process is not affected.
For future devices, it is also possible to separate the power supply and galvanically isolate the communication to reduce the possibilities of hardware attacks and failures.

In this paper, we have also shown the feasibility of our robust architecture by a \ac{PoC} implementation on a Cortex\textsuperscript{\tiny{\textregistered}}-M7 \ac{MCU} for the network tasks combined with a Cortex\textsuperscript{\tiny{\textregistered}}-M0 \ac{MCU} for the time-critical \ac{IO} handling.
The network \acp{MCU} runs FreeRTOS and the \ac{IO} MCU runs a bare metal system.
They communicate over \ac{SPI} with each other in such a way that the timing behavior is predictable.
Our benchmark experiments have shown that physical controlling can be influenced by a deviation maximum of the cycle time of under one percent.
These experiments were performed during a simulated \ac{DoS} flooding attack.
The results show that our dual \ac{MCU} approach is a feasible solution against these kinds of network attacks on \ac{IIoT} devices such as \acp{PLC}.



\bibliographystyle{ieeetr}
\bibliography{\jobname}


\end{document}